\documentstyle[12pt]{article}

\def\be{\begin{equation}}
\def\ee{\end{equation}}
\def\a{\alpha}
\def\e{\epsilon}

\def\d{\partial}
\def\arctg{\mbox{arctg}}
\def\cos{\mbox{cos}}
\def\tg{\mbox{tg}} 
\def\sh{\mbox{sh}}
\def\ch{\mbox{ch}}
\def\tp{t^{\prime}}
\def\tpp{t^{\prime\prime}}

\def\g{\gamma}
\def\w{\omega} 
\def\sign{\mbox{sign}}

\begin{document}

\begin{center}
{\bf The interaction energy of the particle-hole pair in the XXZ- spin chain.}
\end{center}
\vspace{0.2in}
\begin{center}
{\large A.A.Ovchinnikov}
\end{center}

\begin{center}
{\it Institute for Nuclear Research, RAS, Moscow}
\end{center}

\vspace{0.2in}

\begin{abstract}

We calculate directly the interaction energy of the eigenstate with 
the single particle and the single hole for the XXZ- spin chain.  
We find the connection of this addition to the energy with the 
two- particle forward scattering phase shift function. 
The results can have applications to the calculation of the exponents 
of the threshold singularities in the correlators of the XXZ- spin chain.

\end{abstract}

\vspace{0.3in}

{\bf 1. Introduction}

\vspace{0.2in}

It is well known how to calculate the energy of the eigenstates with 
the single particle or the single hole in the infinite volume limit 
in the integrable models (for example see \cite{G}). However the 
energy of the eigenstate with two such excitations is not the sum of 
two excitation energies but have the additional term of order $1/L$, 
where $L$ is the size of the system, which we call the interaction 
energy. It is the goal of the present Letter to calculate the 
particle-hole interaction energy for the XXZ- spin chain. 
The magnitude of the particle-hole interaction energy is important 
for many applications among which is the calculation of the exponents 
of the threshold singularities in the correlators of the XXZ- spin chain. 
For example the knowledge of the interaction energy allows one to 
obtain the parameters of the mobile impurity effective Hamiltonians 
(for example, see \cite{Gl},\cite{RMP})  
responsible for the threshold singularities in a simple and unambiguous 
way. Our predictions for the parameters of this effective Hamiltonian 
($V_L$ and $V_R$ in the notations of ref.\cite{Gl}) 
are different from the results of \cite{PWA},\cite{CP},
which is connected with the shift of the dressed energy due to the shift 
of the rapidity of the high energy particle.  
Although the method of calculation based on the effective mobile impurity 
Hamiltonian \cite{PWA},\cite{CP} presumably gives wrong results for the 
exponents of the threshold singularities for the 
XXZ model, we believe that the correct method will also use 
the knowledge of the particle-hole interaction energy. 
The same method can be used to calculate the forward scattering phase 
shifts and the interaction energies for the eigenstates with 
two particles and two holes. 
Let us make the definition of the interaction energy more precise. 
We consider the system on the finite interval of length $L$. Then 
up to the terms of order $\sim 1/L$ the energy of the eigenstate 
with the particle (rapidity $t_1$) and the hole (rapidity $t_0$) 
has the form: 
\[
E=E_0+\e_L(t_1)+\e_L(t_0)+\Delta E(t_1,t_0). 
\]
Here $\e_L(t_1)$ and $\e_L(t_0)$ are the excitation energies of the 
single particle and the single hole at finite $L$. 
At $L\rightarrow\infty$ their form is well known and we did not 
calculate the corrections $\sim 1/L$ to these energies since this 
requires the special techniques. Instead we calculate the quantity 
$\Delta E$ which we call the interaction energy and which is also of order 
$1/L$. The separation of different terms of order $\sim 1/L$ will be 
performed automatically in the process of calculations. 
In Section 2 we calculate the particle-hole forward scattering phase shift 
for the XXZ- spin chain and present the heuristic arguments in favor of 
its connection with the particle-hole interaction energy. In Section 3 
we calculate the particle-hole interaction energy directly and show that 
in fact it is related to the forward scattering phase shift.

\vspace{0.3in}

{\bf 2. Particle-hole scattering phase shift. }

\vspace{0.2in}

Let us calculate the two-particle particle-hole forward scattering 
phase shift for the XXZ- model following the method of Korepin \cite{K}. 
The Hamiltonian of the model has the form: 
\[
H=\sum_{i=1}^{L}(S_{i}^{x}S_{i+1}^{x}+S_{i}^{y}S_{i+1}^{y}
+\Delta S_{i}^{z}S_{i+1}^{z}), 
\]
where the periodic boundary conditions are assumed, $L$ is the 
length of the chain and the anisotropy parameter $\Delta=\cos(\eta)$. 
The Bethe Ansatz equations for the roots (rapidities) 
$t_{\a}$, $\a=1,\ldots M$ can be represented in the following form: 
\be
L\phi(t_{\a})=2\pi n_{\a}+\sum_{\g\neq\a}\phi_2(t_{\a}-t_{\g}), 
\label{n}
\ee
where $n_{\a}$ are the integer (half-integer) quantum numbers 
and the functions $\phi(t)$, $\phi_2(t)$ are given by 
\[
\phi(t)=\frac{1}{i}\ln\left(-\frac{\sh(t-i\eta/2)}{\sh(t+i\eta/2)}\right),~~
\phi_{2}(t)=\frac{1}{i}\ln\left(-\frac{\sh(t-i\eta)}{\sh(t+i\eta)}\right). 
\]
The energy of the eigenstate is equal to $E=-(\sin(\eta)/2)C$, where 
\[
C=\sum_{\a=1}^{M}C(t_{\a}), ~~~~~C(t)=\phi^{\prime}(t). 
\]
Let us denote by $t_1$ and $t_0$ the roots corresponding to the particle and 
the hole respectively. For the particle excitation $t_1=\tp_1+i\pi/2$ 
is the complex number ($\tp_1$ is real, for example see \cite{O}). 
Let us denote by $\theta_{1}(t_1)$ the phase which is picked up during the 
process of transfer of the particle $1$ around the circle of length $L$ for 
the component of the Bethe wave function without the hole proportional to 
$\exp(ik_{\a}x_{\a})$ without the permutations ($x_{\a}$, $k_{\a}$ are the 
coordinates and the momenta). Let us denote by $\theta_{2}(t_1)$ the same phase 
in the presence of the hole with the rapidity $t_0$. Then according to \cite{K} 
the dressed phase of scattering of the particle $t_1$ on the hole $t_0$ is given 
by the equation: 
\be
\Delta\theta(t_1-t_0)=\theta_{1}(t_1)-\theta_{2}(t_1)= 
-\phi_2(t_1-t_0)-\sum_{\g\neq 1}\phi_2^{\prime}(t_1-t_{\g})(\tp_{\g}-t_{\g}), 
\label{dt}
\ee
where $t_{\a}$ are the solutions of Bethe equations (\ref{n}) with particle and 
without the hole and $\tp_{\a}$ are the roots in the presence of the hole. 
Using the standard procedure introducing the function 
$W(t_{\a})=L(\tp_{\a}-t_{\a})$ and denoting by $R(t)$ the density of roots we get 
for the function $RW(t)=R(t)W(t)$ the equation: 
\[
2\pi RW(t)+\int d\tp\phi_2^{\prime}(t-\tp)RW(\tp)= 
-\phi_2(t-t_0). 
\]
The solution of this equation is $RW(t)=-\tilde{F}(t-t_0)$, where the function 
$\tilde{F}(t)$ is given by the equations 
\be
\tilde{F}^{\prime}(t)=F(t), ~~~~
F(\w)=\int dte^{i\w t}F(t)=\frac{\phi_2^{\prime}(\w)}{2\pi+\phi_2^{\prime}(\w)}, 
\label{F}
\ee
where $\phi_2^{\prime}(\w)=\int dte^{i\w t}\phi_2^{\prime}(t)$. 
Substituting this solution into the equation (\ref{dt}) we obtain in  
thermodynamic limit the equation for the phase: 
\be
\Delta\theta(t_1-t_0)=-\phi_2(t_1-t_0)+
\int dt\phi_2^{\prime}(t_1-t)\tilde{F}(t-t_0). 
\label{phase}
\ee
In order to evaluate the phase (\ref{phase}) we first take the derivative 
of both sides of this equation and then make use of the Fourier transform. 
The result of the calculations has the form: 
\be
\Delta\theta^{\prime}(\w)=
-2\pi\frac{\tilde{\phi}_2^{\prime}(\w)}{2\pi+\phi_2^{\prime}(\w)}, 
\label{tomega}
\ee
where we denote 
\[
\Delta\theta^{\prime}(\w)=\int dte^{i\w t}\Delta\theta^{\prime}(t), ~~~~
\tilde{\phi}_2^{\prime}(\w)=\int dte^{i\w t}\phi_2^{\prime}(t+i\pi/2). 
\]
Note that these functions are defined as the Fourier transforms of the 
functions with real arguments after the shift $i\pi/2$ in 
$t_1=t_1^{\prime}+i\pi/2$ ($t_1^{\prime}$- is real) is taken into account. 
This is indicated by the tilde in $\tilde{\phi}_2^{\prime}(\w)$ and 
the function $\Delta\theta^{\prime}(\w)$ corresponds to the Fourier 
transform of the function $\Delta\theta^{\prime}(t_1^{\prime}-t_0)$ 
on the real axis throughout the paper.  
Note also that the function $\tilde{\phi}_2(t)=\phi_2(t+i\pi/2)$ is not 
continuous, it has a jump of the magnitude $2\pi$ at $t=0$. However 
one can define the continuous function $\tilde{\phi}_2(t)_{reg}$ 
without the jump, which leads to the following relation 
$\tilde{\phi}_2^{\prime}(\w)=2\pi+\tilde{\phi}_2^{\prime}(\w)_{reg}$. 
Now one can easily obtain the Fourier transforms: 
\be
\phi_2^{\prime}(\w)=2\pi\frac{\sh(\w\pi/2-\w\eta)}{\sh(\w\pi/2)}, ~~~
\tilde{\phi}_2^{\prime}(\w)_{reg}=-2\pi\frac{\sh(\w\eta)}{\sh(\w\pi/2)}. 
\label{Fourier}
\ee
Thus from the equation (\ref{tomega}) we obtain the following final 
result for the phase shift: 
\be
\Delta\theta^{\prime}(\w)=
-\pi\frac{\sh(\w\pi/2)-\sh(\w\eta)}{\sh(\w\pi/2-\w\eta/2)\ch(\w\eta/2)}.
\label{theta}
\ee
From the equation (\ref{theta}) one can see that the function 
$\Delta\theta(t)$ has a jump of the magnitude $-2\pi$ at $t=0$ such that 
$\Delta\theta(-0)=\pi$, $\Delta\theta(+0)=-\pi$. One can also find this 
function at infinity using the equation:
\[
\Delta\theta^{\prime}(\w=0)=\int dt\Delta\theta^{\prime}(t)=
\Delta\theta(\infty)-\Delta\theta(-\infty). 
\]
Thus we obtain the value 
\[
\Delta\theta(\pm\infty)=\mp\pi(1-1/\xi), ~~~~\xi=2\frac{\pi-\eta}{\pi}, 
\]
where $\xi$- is the usual Luttinger liquid parameter for the XXZ- 
spin chain. One can also determine the value of the derivative 
$\Delta\theta^{\prime}(t=+0)$. We will not reproduce the corresponding 
integral here. 
Of course in the limit of the XX- spin chain $\eta=\pi/2$ we obtain 
$\Delta\theta=0$ and zero interaction energy as it should be for the 
Free-Fermion point. 
Concluding these calculations let us present the simple 
expression for the phase shift which follows from the equality 
$\Delta\theta(\w)=-2\pi\tilde{\phi}_2(\w)/(2\pi+\phi_2^{\prime}(\w))$. 
In fact we have 
\be
\Delta\theta(t_1-t_0)=-2\pi\tilde{F}(t_1-t_0), 
\label{Ftheta}
\ee
where the function $\tilde{F}(t)$ was introduced in (\ref{F}).

Now from the heuristic arguments one can determine the interaction energy 
as a function of variable $t_1^{\prime}-t_0$. 
The total effective Hamiltonian which correctly reproduce the terms of 
order up to $1/L$, where $L$ is the length of the chain, has the form 
$H_{eff}=\e(p_1-i\d_1)+\e(p_2-i\d_2)+U\delta(x)$. 
Subtracting the constant term $\e(p_1)+\e(p_2)$, we obtain the following 
effective Hamiltonian:  
\be
H_{eff}=-i(v_1\d_1+v_2\d_2)+U\delta(x), 
\label{eff1}
\ee
where $\d_i=\d/\d x_i$ and $x=x_{12}=x_1-x_2$. 
Here $x_1$ is the coordinate of the particle and $x_2$ is the 
coordinate of the hole. In eq.(\ref{eff1}) we have $v_{1,2}=v\cos(p_{1,2})$ 
and the coupling constant $U$ depending on the momenta $p_1$ and $p_2$. 
Substituting the wave function of relative motion of two particles 
\[
\psi(x)=\theta(-x)e^{i\Delta px/2}+\theta(x)e^{i\Delta px/2+i\delta}, 
\]
where $\Delta p=p_1-p_2$, into the Schrodinger equation for the Hamiltonian 
(\ref{eff1}) after the integration over the small vicinity of the point $x=0$ 
we obtain the following expression for the phase $\delta$: 
\be
\delta=2\arctg(U/2\Delta v), 
\label{delta}
\ee
where $\Delta v=v_1-v_2$. If the system is considered on the finite interval of 
length $L$ from the equations for the momenta $p_{1,2}$ of the form
\[
p_1=q_1-\delta/L, ~~~~p_2=q_2+\delta/L, ~~~~q_{1,2}=2\pi n_{1,2}/L, 
\]
we obtain the energy of the form 
$E=v_1p_1+v_2p_2=v_1q_1+v_2q_2-\Delta v\delta/L$. 
Thus after the identification of $\delta$ with the phase $-\Delta\theta$ we 
obtain the interaction energy of the particle-hole pair for the 
$XXZ$ Hamiltonian 
\be
\Delta E=\frac{\Delta v\Delta\theta}{L}.
\label{de}
\ee
In the next Section we will reproduce the result (\ref{de}) 
by direct calculations. 

The effective Hamiltonian (\ref{eff1}) is not valid for the eigenstates 
with two particles or two holes since they obey the Fermi- statistics. 
However the corresponding scattering phase shifts can be introduced 
in a similar way. For example, for the hole-hole eigenstate the 
scattering phase shift $\Delta\theta_{hh}(t_{01}-t_{02})$, where 
$t_{01}$ and $t_{02}$ are the rapidities of the holes, is given by the 
equation: 
\[
\Delta\theta_{hh}^{\prime}(\w)=-2\pi F(\w), ~~~
F(\w)=\frac{\phi_2^{\prime}(\w)}{2\pi+\phi_2^{\prime}(\w)}= 
\frac{\sh(\w\pi/2-\w\eta)}{2\sh(\w\pi/2-\w\eta/2)\ch(\w\eta/2)}. 
\]
Now the function $\Delta\theta_{hh}(t)$ is the smooth function at $t=0$. 
The particle-particle phase shift is given by the similar formula.

\vspace{0.2in}

{\bf 3. Direct calculation of the particle-hole interaction.}

\vspace{0.2in}

In order to determine the interaction energy by the direct 
calculations we will make use of the following simple trick. 
Instead of considering the single hole at the rapidity $t_0$, 
imagine that we have the $\Delta N_0$ holes located around $t_0$. 
Assuming that the density $n_0=\Delta N_0/L$ is small it is clear 
that the terms of order $n_0$ in the expansion of the energy 
will reproduce correctly the leading $O(1/L)$ correction to the 
energy in the case of the single hole 
(formally one can substitute $n_0=1/L$ in the final expression). 
This corresponds to the term $\e_L(t_0)$ in the expansion of the 
energy presented in the Introduction. 
Note that the interaction energy is also of order $\sim n_0$ 
which allows one to calculate $\Delta E$ without the calculation 
of $\e_L(t)$. 
We calculate the response of the system with holes to adding 
an extra particle with the rapidity $t_1$. 
The addition to the energy have two sources. First, the roots 
on the real axis are shifted by the magnitude of order $\sim 1/L$. 
Second, for a given quantum number $n_1$ the value of the rapidity 
$t_1$ is shifted due to the presence of the holes. 
Of course, the dressed energy of the additional particle $t_1$ in 
the absence of the holes should be subtracted. In principle the 
calculations can be performed for arbitrary $\Delta N_0$. 
Clearly, in this way we will take into account all the possible 
contributions to the shift of energy. 
Consider the first source. The calculations are quite standard 
and we will omit the details.  
We start with the change of energy due to the additional particle 
and consider the contribution of real roots (here we use the 
normalization of the energy $E=-(\sin(\eta)/2)C$): 
\be
\Delta C_{(+1)}=C(t_1)+\sum_{\a\in\Lambda}\left(C(t_{\a}^{\prime})
-C(t_{\a})\right), ~~~C(t)=\phi^{\prime}(t), 
\label{sum}
\ee
where $\Lambda$ denotes the region of the real axis occupied 
by the particles (roots). After some simple algebra we obtain 
$\Delta C_{(+1)}=-2\pi R_0(t_1)+\Delta C_1$ 
where the first term is the dressed energy of the particle in the 
infinite volume limit (without the hole). 
Then the final result for the interaction energy due to 
the shift of the real roots has the form: 
\be 
\Delta C_1=-2\pi\int_{O}dtR_0^{\prime}(t)RW(t),  
\label{C1}
\ee
where $R_0(t)=1/2\eta\ch(\pi t/\eta)$ is the unperturbed 
density of roots, $O$- is the region of the real axis occupied by the 
holes, and the function $RW(t)$ is the solution of the equation 
\be
2\pi RW(t)+\int_{\Lambda}d\tp\phi_2^{\prime}(t-\tp)RW(\tp)=
\phi_2(t-t_1).   
\label{EQ1}
\ee 
The equations (\ref{C1}),(\ref{EQ1}) are valid for an arbitrary 
regions $\Lambda$ and $O$ of the real axis. 
Starting from this point we assume that the region $O$ is small 
($n_0<<1$). 
The calculation of the sum in the equation (\ref{sum}) was 
performed in the thermodynamic limit. Let us briefly comment 
on this point. In principle in the sum (\ref{sum}) both the terms
of order $1/L$ and $n_0=\Delta N_0/L$ are required 
(in order to transform $\e(t_1)$ into $\e_L(t_1)$). 
However it is clear that all the terms of order $\sim n_0$ 
are the interaction energy $\Delta E$, which is evidently 
$\sim n_0$. Thus we can neglect the terms of order $1/L$ 
and retain the terms of order $\sim n_0$. So for our purpose 
one can use the expressions (\ref{C1}), (\ref{EQ1}) obtained 
in the thermodynamic limit and single out the terms $\sim n_0$ 
in the expression for the energy. 
This will automatically give us the interaction energy $\Delta E$. 
Thus from the equations (\ref{C1}),(\ref{EQ1}) one can easily find 
that to the leading order in $n_0$ the result is 
\be
\Delta C_1=2\pi(R_0^{\prime}(t_0)/R_0(t_0))\tilde{F}(t_1-t_0)n_0. 
\label{Res1}
\ee
Evaluating the coefficient before $\tilde{F}$ in (\ref{Res1}) and 
substituting $n_0\rightarrow 1/L$, we get exactly the energy 
\be
\Delta E_1=-\frac{v_0\Delta\theta(t_1-t_0)}{L}, 
\label{final1}
\ee
where $v_0$ is the velocity corresponding to the rapidity $t_0$. 
Now let us consider the change of the energy due to the shift  
of the root $t_1$. Here one should evaluate the difference 
$\epsilon(\tpp_1)-\epsilon(t_1)$, where $\epsilon(t)$ is the 
dressed energy of the particle with the rapidity $t$ in the 
infinite volume limit. Up to the standard normalization constant 
$-\sin(\eta)/2$ this function equals $\epsilon(t)=-2\pi R_0(t)$. 
Thus in the leading order in $n_0$ we have 
\[
\Delta C_2=-2\pi R_0^{\prime}(t_1)(\tpp_1-t_1).
\]
The expression for the difference of roots in this equation 
can be found from the Bethe Ansatz equations and has the form: 
\be
-L2\pi(\tpp_1-t_1)R_0(t_1)
+\int_{\Lambda}dt\phi_2^{\prime}(t_1-t)RW(t)=
-\Delta N_0\phi_2(t_1-t_0),  
\label{diff}
\ee
where the function $RW(t)$ obeys the equation 
\[
2\pi RW(t)+\int d\tp\phi_2^{\prime}(t-\tp)RW(\tp)= 
-\Delta N_0\phi_2(t-t_0), 
\]
which has the solution $RW(t)= -\Delta N_0\tilde{F}(t-t_0)$. 
Substituting this function into the equation (\ref{diff}) we obtain: 
\be
- L2\pi(\tpp_1-t_1)R_0(t_1)=\Delta N_0\left[-\phi_2(t_1-t_0)+ 
\int_{\Lambda}dt\phi_2^{\prime}(t_1-t)\tilde{F}(t-t_0)\right].
\label{extra}
\ee
One can see that 
in the leading order in $n_0$ the expression in the parenthesis 
in this equation is nothing else but the phase shift function 
$\Delta\theta(t_1-t_0)$.  
Note that here we really write $R_0(t_1)$ instead of $R_0(\tp_1)$ 
for simplicity throughout the paper 
which leads to an extra minus sign in 
(\ref{diff}),(\ref{extra}) (for example, see Appendix A). 
Thus combining all factors we obtain the shift of the energy due
to the shift of the root $t_1$ in the form: 
\be 
\Delta E_2=\frac{v_1\Delta\theta(t_1-t_0)}{L}. 
\label{final2} 
\ee
In principle we could calculate this contribution considering 
the energy $\e(t_0^{\prime})$, where the value $t_0^{\prime}$ 
corresponds to the shift of $t_0$ due to the presence of the 
particle. However in this case it is not obvious that 
we will obtain all possible contributions to the interaction 
energy. 
Then the sum of the contributions (\ref{final1}) and 
(\ref{final2}) is the total interaction energy 
\be
\Delta E=\Delta E_1+\Delta E_2=
\frac{(v_1-v_0)\Delta\theta(t_1-t_0)}{L}, 
\label{total}
\ee
in agreement with the expression (\ref{de}). 
The equation (\ref{total}) is the final result of the present 
Letter. It relates the particle-hole interaction energy 
with the two-particle scattering phase shift. 
Note that the energy (\ref{total}) can be expressed through 
the momenta of the particle and the hole (see the Appendix A). 
At small difference of the momenta defined with respect 
to the Fermi- momentum we reproduce the equation (\ref{delta}).
Let us note that using the same method one can calculate the 
energy for the particle-particle and the hole-hole interaction. 
For example, for two holes the interaction energy has the form: 
\[
\Delta E_{hh}=
-\frac{(v_{01}-v_{02})\Delta\theta_{hh}(t_{01}-t_{02})}{L}, 
\]
where $v_{01}$ and $v_{02}$ are the velocities of the holes. 
The modification of our method is required 
in the case when the region $O$ is at the right end of the 
real axis and the formal substitution $t_0\rightarrow\infty$ 
in (\ref{total}) does not give the correct result for the 
hole at the Fermi- momentum. The reason is that we have assumed 
that $R(t)\simeq R_0(t)$ in the region $O$, which is obviously 
correct when this region is far enough from the infinity. 
However this is not correct when the region $O$ is at the right 
end of the real axis which leads to the different result 
for the energy in the case when the holes are located 
near the right (left) Fermi- point (see Appendix B). 
Since we are interested in the $1/L$ corrections, we should 
always keep $L$ finite. Then the transition from one regime 
to another should occur at $t_0\sim\ln L$, which is of order 
of the maximal root in the system without the excitations. 
The detailed analysis of this transition is beyond the scope 
of the present Letter.   
Thus from the results of the Appendix B one can easily obtain 
the parameters of the effective Hamiltonian \cite{Gl}, 
which does not coinside with the predictions of 
\cite{PWA},\cite{CP}. 
It is not the goal of the present Letter to discuss the  
correct way to evaluate the threshold singularities 
for the XXZ- spin chain.

\vspace{0.2in}

{\bf 4. Conclusion.}

\vspace{0.2in}

In conclusion, we did not calculated all the terms of order 
$1/L$ in the expansion of energy for the eigenstate with the 
single particle and the single hole. It is a separate problem 
to calculate the energy of the single particle up to the terms 
of order $1/L$. These calculations were performed for example 
in \cite{PWA1} using completely different technique. 
At the same time the application of the method presented 
here to the calculation of $\e_L(t)$ gives the result 
$\e_L(t)=\e(t)+O(1/L^2)$, so that the interaction energy 
is the total $1/L$- correction to the energy. 
At present time we dod not know where is the mistake 
in the results of ref.\cite{PWA1}, so we think that now 
the total value of the $O(1/L)$ corrections to the energy 
is an open question. 
What we have presented in the present Letter 
is the calculation of the interaction energy i.e. of the 
excess of energy of a particle in the presence of a hole. 
It is exactly this energy which is used for example in the 
mobile impurity effective Hamiltonian responsible for the 
threshold singularities of the correlators in different models 
(for example, see \cite{RMP} and references therein). 
The reason why this method does not give the universal phase 
shifts of the threshold singularities 
at zero momentum is that the spin operators in terms 
of the mobile impurity operator in the effective model does 
not have a simple form. 
However we believe that our results can be useful for this problem.

\vspace{0.2in}

{\bf Appendix A.}

\vspace{0.2in}

Here we present the calculation of the momentum of the particle $p_1$ 
as a function of its complex rapidity $t_1=\tp_1+i\pi/2$. 
Considering the Bethe Ansatz equation for the root $t_1$ we obtain 
in the standard way the integral equations for the functions 
$\e(\tp_1)$, $p_1(\tp_1)$, which are dressed energy and momentum 
of the particle corresponding to the root $t_1$. 
After the Fourier transform the solutions of these equations takes 
the form: 
\be
\e(\w)=p_1^{\prime}(\w)=\tilde{\phi}^{\prime}(\w)- 
R_0(\w)\tilde{\phi}_2^{\prime}(\w), 
\label{pe}
\ee
where the functions at the left hand side are related to the functions 
$\e(\tp_1)$, $p_1(\tp_1)$, and the prime means the derivative. 
Taking into account the equations (\ref{Fourier}) and substituting 
$R_0(\w)=1/2\ch(\w\eta/2)$ the equations (\ref{pe}) can be represented 
in the form:
\be
\e(\w)=p_1^{\prime}(\w)=2\pi-2\pi R_0(\w). 
\label{pe1}
\ee 
Let us find the function $p_1(\tp_1)$ from the equation (\ref{pe1}). 
It is the function with the jump at $t=0$ of the form: 
\be
p_1(t)=\pi\sign(t)-2\arctg(e^{\pi t/\eta})+C, 
\label{p2} 
\ee 
where $C$- is some constant which should be fixed from the boundary 
conditions. Thus from (\ref{p2}) we finally obtain the following 
equation for the dependence $p_1(\tp_1)$: 
\be
\tg(-p_1/2+\pi/4+(\pi/2)\sign(\tp_1))=e^{\pi\tp_1/\eta}. 
\label{pfin}
\ee
For comparison the dependence of the momentum on the rapidity for the hole 
has the form:
\be
\tg(p_0/2+\pi/4)=e^{\pi t_0/\eta}
\label{hfin} 
\ee
where $t_0$ and $p_0$ are the rapidity and the momentum for the hole. 
Note that in (\ref{pfin}), (\ref{hfin}) the momenta are defined as the 
total momenta of the eigenstates (not relative to the Fermi- momentum). 
The equations (\ref{pfin}), (\ref{hfin}) allow one to express the 
interaction energy through the momenta of the particle and the hole 
rather than through the difference of the rapidities.  
From the function $\e(\w)$ given by (\ref{pe}), (\ref{pe1}) one can 
also calculate the dressed energy of the particle which is given by 
the same formula as for the hole, $\e(p)=v\sin(p)$, 
$v=(\pi/2)(\sin\eta/\eta)$, where the momentum $p$ is now defined 
with respect to the Fermi- momentum.

\vspace{0.2in}

{\bf Appendix B.}

\vspace{0.2in}

Here we calculate the value of the interaction energy for 
the holes at the Fermi- points with the help of our method. 
In this case the equations (\ref{C1}), (\ref{EQ1}) remain the 
same except that now the region $O$ is located at both ends 
of the real axis, $O=(-\infty;-\Lambda_2)\cup(\Lambda_1;\infty)$, 
$\Lambda=(-\Lambda_2;\Lambda_1)$. Let us solve the equation 
(\ref{EQ1}) first. Using the standard technique (for example, 
see \cite{V}), considering the right Fermi- point, we obtain 
for the function $\chi(t)=RW(t+\Lambda_1)$ the equation
\be
\chi(t)=f(t)+\int_0^{\infty}d\tp F(t-\tp)\chi(\tp), 
\label{equation}
\ee
where the function $f(t)$ equals 
$f(t)=\tilde{F}(t+\Lambda_1-\tp_1)$. 
The solution of the equation (\ref{equation}) has the form 
\be
\chi^{+}(\w)=G^{+}(\w)\int\frac{d\w^{\prime}}{2\pi i}
\frac{1}{(\w^{\prime}-\w-i0)}G^{-}(\w^{\prime})f(\w^{\prime}), 
\label{solution}
\ee
where $\chi^{+}(\w)=\int_0^{\infty}dte^{i\w t}\chi(t)$ and 
the functions $G^{\pm}(\w)$ are the holomorphic functions 
at the upper (lower) half- plane defined by the equation 
\[
F(\w)=1-\frac{1}{G^{+}(\w)G^{-}(\w)}. 
\]
Substituting the Fourier- transform $f(\w)$ into the equation 
(\ref{solution}) and considering the integral over the contour 
in the lower half- plane of the complex variable $\w^{\prime}$ 
one can see that only the residue at $\w^{\prime}=0$ leads to the 
leading order term in the variable $e^{-\pi\Lambda_1/\eta}$. 
Thus we obtain the sulution of the equation (\ref{equation}) 
in the form: 
\be
\chi^{+}(\w)=
\frac{G^{+}(\w)}{\w}\left(\frac{i}{2}(1-1/\xi)\sqrt{\xi}\right). 
\label{sol}
\ee
For the right Fermi- point the expression for the energy takes 
the form: 
\[
\Delta C_1=(2\pi^2/\eta^2)e^{-\pi\Lambda_1/\eta}\chi^{+}(i\pi/\eta). 
\]
Thus using the solution (\ref{sol}) we finally obtain for the energy 
\be
\Delta C_1=(\pi/\eta)\sqrt{\xi}(1-1/\xi)A, 
\label{energy}
\ee 
where the constant $A$ defined according to 
$A=G^{+}(i\pi/\eta)e^{-\pi\Lambda_1/\eta}$ is connected with the 
number of holes at the right Fermi- point. To find this constant 
one has to solve the equation for the density of roots $R(t)$ 
defined at the interval $(-\Lambda_2;\Lambda_1)$: 
\be
2\pi R(t)+
\int_{-\Lambda_2}^{\Lambda_1}d\tp\phi_2^{\prime}(t-\tp)R(\tp)=
\phi^{\prime}(t). 
\label{R}
\ee
The solution of the equation (\ref{R}) for the function 
$\chi(t)=R(t+\Lambda_1)$ at the right end of the real axis has exactly 
the form (\ref{solution}) where the function $f(t)$ now equals 
$f(t)=R_0(t+\Lambda_1)$. Thus we obtain the expression: 
\be
\chi^{+}(\w)=\frac{G^{+}(\w)}{\w+i\pi/\eta}(i/\eta)A, 
\label{solR}
\ee
where the constant $A$ is the same as in the equation (\ref{energy}). 
The value $\chi^{+}(0)=\int_{\Lambda_1}^{\infty}dtR(t)=(\sqrt{\xi}/\pi)A$ 
is connected with the number of holes at the right Fermi- point $n_0$  
and can be easily obtained from (\ref{R}). 
Thus taking into account the normalization factors we finally obtain the 
contribution 
\be
\Delta E_1=-\left(v(1-1/\xi)\pi\xi\right)n_0,  
\label{DE}
\ee
for the right Fermi- point and the same expression (\ref{DE}) for the 
left Fermi- point. 
The expression (\ref{DE}) should be compared with (\ref{final1}). 
One can see that here the extra factor $\xi$ appears. 
The energy (\ref{DE}) taken alone coinsides with the predictions 
of ref.\cite{PWA}. 

Now let us calculate the shift of energy due to the shift of the root 
$t_1\rightarrow \tpp_1$ in the presense of the holes at the edges of 
the real axis. From the Bethe Ansatz equations for this root we find 
for the difference: 
\be
2\pi LR_0(t_1)(\tpp_1-t_1)=\int_{\Lambda}dt\phi_2^{\prime}(t_1-t)RW(t) 
+\int_{O}dt\phi_2(t_1-t)R_0(t), 
\label{1}
\ee 
where the function $RW(t)$ satisfies the equation 
\be
2\pi RW(t)+\int_{\Lambda}d\tp\phi_2^{\prime}(t-\tp)RW(\tp)= 
-\int_{O}d\tp\phi_2(t-\tp)R_0(\tp). 
\label{2}
\ee
Let us stress once more that in (\ref{1}) we use the simplified 
notation $R_0(t_1)$ for the function $R_0(\tp_1)$. 
Combining the equations (\ref{1}),(\ref{2}) we find 
$LR_0(t_1)(\tpp_1-t_1)=-RW(t_1)$.  The contribution to the energy 
equals $\Delta C_2=-2\pi R_0^{\prime}(t_1)(\tpp_1-t_1)$. 
Solving the equation (\ref{2}) in the regime $\tp_1<<\Lambda_1$ 
and neglecting the contributions of the higher order in $n_0$ 
we finally obtain for the contribution to the energy 
of the right Fermi- point: 
\be 
\Delta E_2=v_1\pi(1-1/\xi)n_0, 
\label{3}
\ee
where $v_1$ is the velocity corresponding to the particle $t_1$. 
The contribution of the left Fermi- point is given by the same 
formula (\ref{3}) with an extra minus sign in agreement with the 
equation  (\ref{total}) at $t_0\rightarrow\pm\infty$. 
Thus the interaction energy for the holes located at the Fermi- 
points is found.

\end{document}